\documentclass{sig-alternate}

\usepackage[latin1]{inputenc} 
\usepackage[T1]{fontenc}      
\usepackage{url}

\begin{document}

\title{A Hardware Time Manager Implementation for the {\ttlit Xenomai} Real-Time Kernel of Embedded Linux}
%
%
%
%
%

\numberofauthors{2} 
%
\author{
%
%
\alignauthor
Pierre Olivier\\
       \affaddr{Universit\'e Europ\'eenne de Bretagne, France}\\
       \affaddr{Universit\'e de Brest; CNRS, UMR 3192 Lab-STICC, }\\
       \affaddr{20 avenue Le Gorgeu, 29285 Brest, France}\\
       \email{pierre.olivier@univ-brest.fr}
\alignauthor
Jalil Boukhobza\\
       \affaddr{Universit\'e Europ\'eenne de Bretagne, France}\\
       \affaddr{Universit\'e de Brest; CNRS, UMR 3192 Lab-STICC, }\\
       \affaddr{20 avenue Le Gorgeu, 29285 Brest, France}\\
       \email{boukhobza@univ-brest.fr}
}
\date{}

\maketitle

\begin{abstract}
Nowadays, the use of embedded operating systems in different embedded projects is subject to a tremendous growth. Embedded Linux is becoming one of those most popular EOSs due to its modularity, efficiency, reliability, and cost. One way to make it hard real-time is to include a real-time kernel like Xenomai.
One of the key characteristics of a Real-Time Operating System (RTOS) is its ability to meet execution time deadlines deterministically. So, the more precise and flexible the time management can be, the better it can handle efficiently the determinism for different embedded applications. RTOS time precision is characterized by a specific periodic interrupt service controlled by a software time manager. The smaller the period of the interrupt, the better the precision of the RTOS, the more it overloads the CPU, and though reduces the overall efficiency of the RTOS.
In this paper, we propose to drastically reduce these overheads by migrating the time management service of Xenomai into a configurable hardware component to relieve the CPU. The hardware component is implemented in a Field Programmable Gate Array coupled to the CPU.
This work was achieved in a Master degree project where students could apprehend many fields of embedded systems: RTOS programming, hardware design, performance evaluation, etc.

\end{abstract}

\category{D.4.8}{Operating Systems}{Performances}
\category{C.3}{Special Purpose and Application-based Systems}{Real-time and Embedded Systems}
\category{B.6.1}{Logic Design}{Design Style}[logic arrays]
\category{K.3.2}{Computers and Education}{Computer and Information Science Education}

\keywords{Real-Time Operating Systems, Embedded Linux, Reconfigurable Architectures, FPGA, Embedded Systems} 

\section{Introduction}

Nowadays Real-Time Operating Systems (RTOS) are integrated into many embedded devices. Using an RTOS provides many benefits. First, by abstracting the hardware layer, the OS facilitates the programmers work, and the standardization of the software layer. Using an RTOS also brings an executive software platform, providing some important \textit{services} : 1) \textit{Task scheduling} ; 2) \textit{Time management} : giving the RTOS the notion of time ; 3) \textit{Inter-task communication and synchronization mechanisms} (IPC), etc. Even though RTOSs provide many advantages, the time they spend managing their own structures can be considered as \textit{overhead} that RTOS designers should minimize.

RTOS are generally executed in a constrained environment. First, the RTOS must deal with traditional embedded constraints : limited hardware resources, especially in terms of memory footprint, generally limited computing power, energy consumption constraints, high reliability, and also fast Time To Market. Moreover, RTOS are subject to specific constraints, which are the \textit{predictability} of their behavior, and the \textit{determinism} of the tasks execution time. 

RTOSs lie upon a hardware timer periodically generating an interrupt notifying the CPU of a generated clock tick. The corresponding handler is executed and calls the tick management function of the RTOS. Depending on the RTOS this function performs various tasks, but one can identify the following main jobs : 1) maintaining a \textbf{system time} variable, counting the number of elapsed clock ticks since the boot of the system ; and 2) notifying different timers that a clock tick has occurred, and performing particular actions for expired timers. Some of those timers are dedicated to manage \textbf{periodic tasks}.
 
A high resolution timer allows the system to be more precise and responsive. Nevertheless, it also means a tick management function executed more frequently, though a higher CPU load. Depending on the load, a weak CPU with a high resolution timer can generate indeterminism in time management and so missed deadlines \cite{Labrosse02}. A compromise is then to be made between performance and time precision. 

In this context, several studies have proposed to migrate some RTOS services from a software toward a hardware implementation. This is performed for two main reasons: 1) A performance improvement, by benefiting from spatial and temporal (pipeline) parallelisms provided by such a hardware implementation. 2) To release the CPU from the corresponding service execution, thus reducing the generated overhead. Such improvements would allow systems designers to use less efficient but less expensive, and power consuming components. 

One of the RTOS services one can need to optimize and/or configure according to the application needs is the time manager. We propose to enhance this service in the Xenomai real-time Linux framework. We present a configurable hardware architecture for a simple time manager implemented on an FPGA circuit, connected to a CPU executing the Xenomai framework in an embedded Linux.

The \textit{tick-less} mode of some OS allows them not to be interrupted at each timer tick, but only when needed. This is done by carefully loading one-shot timers and responding to various hardware interrupts. The work presented in this paper is done to enhance performance of \textit{non-tick-less} RTOSs, which is the case of most EOS. As Xenomai provides both tick-less and non-tick-less mode, we used the non-tick-less mode.

This project was proposed in a second year Master (Software for Embedded Systems at the Universit\'e of Brest - France) course on embedded operating systems. The objective of the project was to cover many domains of embedded systems throughout one project: hardware architecting and programming (FPGA, Hardware Design Languages), real-time systems, embedded operating systems development, device driver and kernel programming and integration, performance evaluation and validation procedures, etc.

In this paper, we first introduce some state-of-the-art studies about hardware implementations of RTOS services. Next, we present the Xenomai real-time framework for embedded Linux. In the third section, we describe the architecture of the hardware time manager, and its integration in the system. We finally give some performance evaluation results before concluding.

\section{State of the Art}
A real-time operating system must be able to respond within a deterministic time. The latency generated by the use of the RTOS (\textit{overhead}) must not interfere with the reactivity of the system. Some studies focus on migrating software services to hardware components in order to unload the CPU, thus reducing the RTOS related overhead thereby enhancing the applicative performance.

We can find many hardware implementations of  the scheduler in the literature. In \cite{Kuacharoen03}, the authors present a configurable hardware scheduler, supporting various scheduling policies : \textit{Priority-based}, \textit{Rate Monotonic}, and \textit{Earliest Deadline First}. The \textit{Spring Scheduling Co-Processor} (\textit{SSCoP}) \cite{Burleson93} is also a hardware implementation of the scheduler. This system shows an improvement factor of 6.5 as compared to the software version.

 \textit{Real-time Task Manager} (\textit{RTM}) \cite{Kohout03} is a component handling scheduling functions, but also time and event management. The authors show that OS ($\mu$C/OS-II, NOS) latency and system response time are considerably enhanced with RTM.

Finally, \textit{Fastchart} \cite{Lindh91} is a complete hardware real-time kernel. To obtain a fully deterministic EOS, the authors remove features such as CPU pipelines and cache, and DMA. As a software operating system running on such hardware is drastically slower, Fastchart is fully implemented in hardware.  

To the best of our knowledge, no study has been realized on the migration of the time manager service of the Xenomai kernel on reconfigurable hardware (FPGA). This allows to have a configurable time manager that can be tuned and dimensioned according to application needs, should the hardware platform contain an FPGA circuit. One of the latest innovation of Intel is  precisely the integration of an FPGA and an Atom processor into the same chip (Intel atom E6x5C series).

\section{The Xenomai real-time frame-\\work for Linux}

In this section we briefly present the Xenomai real-time framework for Linux, and the Adaptive Domain Environment for Operating System (Adeos) layer, which allows Xenomai and Linux to run on the same hardware platform.

\subsection{Adeos layer}

\textit{Adeos} \cite{Yaghmour01} is a resource virtualization layer, allowing multiple entities called \textit{domains}, that can be seen as complete operating systems, to run  simultaneously on the same hardware platform \cite{Gerum05}.

Adeos domains can compete with each other for receiving system generated \textit{events}. Those events can be incoming external (or virtually generated) interruptions, Linux system calls invocations, or various kernel-code-related events like context switches. Adeos introduces the \textit{event pipeline}, which can be seen as a chain of domains of decreasing priority. The events are consequently propagated throughout the pipeline, distributed firstly to the utmost priority domain, then distributed to lower priority domains.

\begin{figure}[h!]
  \centering
  \includegraphics[width=0.45\textwidth]{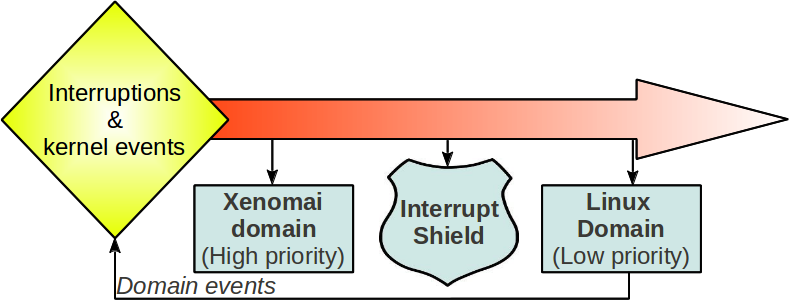}
  \caption{Adeos Event pipeline when running the Xenomai with Linux kernel (adapted from \cite{Gerum05}).}
  \label{EventPipeline}
\end{figure}

In the case of a Xenomai RTOS running with a Linux kernel, the Adeos pipeline and the organization of the domains is depicted in Figure \ref{EventPipeline}. In the pipeline,  we can see that Xenomai has the highest priority, so it can handle and manage first the events before passing them to the Linux kernel. The events can also be blocked by the interrupt shield, preserving the real-time framework from latencies due to event management by the Linux kernel. Throughout this mechanism, Xenomai framework can provide real-time guarantees.

\subsection{Xenomai}

Xenomai \cite{Kiszka05} provides a kernel-based Application Programming Interface (API) for real-time applications. A user-land API is also available, at the cost of longer latencies. Xenomai introduces the concept of \textit{skins}. Skins are source codes emulating proprietary APIs used for porting real-time applications from various RTOSs such as \textit{VxWorks}, \textit{pSOS}, etc. to Xenomai. When designing a Xenomai application from scratch, the \textit{native} Xenomai skin can be used.

All Xenomai skins rely on the \textit{nucleus}, the core of the RTOS, implementing all algorithms for real-time functionalities. Xenomai provides all standard services one can expect to find in a RTOS\footnote{For more information about Xenomai, one can browse the Xenomai website : \url{http://www.xenomai.org}} : task management, multiple scheduling algorithms, IPCs, etc.

\subsection{Time management in Xenomai}
We focused in this project on periodic real-time tasks that make extensive use of timers. A commonly used skeleton for those tasks in Xenomai as follows :

\begin{verbatim}
void myRealTimeTask(void *arg) {
  SetPeriodic(myself, period);
  while(1) {
    /* Do something */
    wait_period();
  }
}
\end{verbatim}

The \verb+SetPeriodic()+ function creates and starts a timer related to the calling task, with a period (in clock ticks) equals to the specified parameter. The global tick management function notifies the timer each time a clock tick occurs. When reaching wake up time, the timer executes a handler placing the task in the ready state.

\section{Architecture and integration of the hardware time manager}

In this section we present the architecture of the proposed hardware time manager component, and the way it is integrated into the Xenomai framework.

From the RTOS point of view, the hardware time manager should provide the following basic time management operations :
\begin{itemize}
  \item \textbf{GetTime} and \textbf{SetTime}: read/modify the value of the system time ;
  \item \textbf{TaskDelay}: load a counter of a delayed task for a given amount of clock ticks ;
  \item \textbf{GetTasksToWake}: obtain the identifiers of tasks that need to be awaken (if any) ;
  \item \textbf{ClearTask}: acknowledge from the CPU indicating the awakening of a task.
\end{itemize}

\subsection{Hardware architecture and integration}

\begin{figure}[h!]
  \centering
  \includegraphics[width=0.4\textwidth]{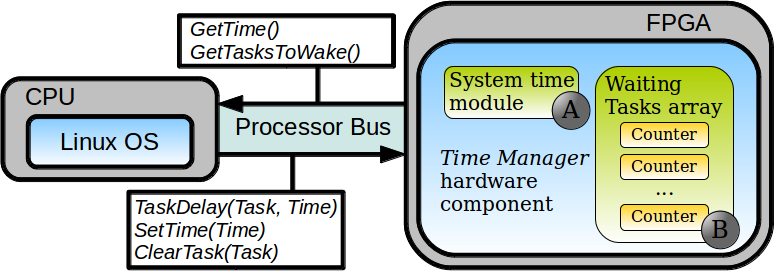}
  \caption{\textbf{Block view of the hardware time manager and its integration in a Linux-based system.}}
  \label{Architecture}
\end{figure}

The designed hardware time manager is composed of two main modules, which represent the two main functions we want to support : 1) maintaining the global system time and 2) allowing tasks to suspend their execution for a specific amount of time. Thus, the two main modules composing the hardware time manager are the system time module and the waiting tasks array as seen in Figure \ref{Architecture}.

The \textit{system time module} is a simple counter which value is initialized to zero when the system starts (i.e. when the bitstream is loaded on the FPGA). This counter is then incremented each FPGA clock cycle.

The \textit{array of waiting task module} is an array of counters. The number of counters it contains is equal to the number of tasks that can potentially be put in a waiting state (there is space for optimization for this module). When a task needs to be put in a waiting state, the corresponding counter is loaded with the appropriate number of ticks. This counter is then decremented at each clock cycle.

All counters in the component are 64 bit width which corresponds to a very high frequency/precision counter. Counter size can be reduced according to the needed  precision.

The architecture was realized in an incremental approach. Students were first provided with some simple hardware component to test, then a very simple counter, and finally the whole module was introduced.

\subsection{Communication between CPU and IP}
Communication between the CPU (executing the operating system) and the hardware time manager is performed through a register-based interface. 

On our evaluation board, the FPGA chip is directly connected to the CPU through a dedicated bus. We defined some control registers  (i.e. FPGA registers) in the hardware part that were mapped to the device driver's process address space. A kernel driver was written in C, as well as a user-space one. The memory mapping is performed using the \verb+mmap()+ system call in user-land, and \verb+ioremap()+ in kernel space. One function is provided for each  basic operations supported by the hardware component.

When the counter corresponding to a waiting task reaches zero, the related task needs to be awakened at the OS level. The component then triggers an (hardware) output signal at a high level and waits for an acknowledgment from the CPU meaning that the task is actually awake. This specific signal is plugged on the interrupt output of the CPU.

Here again, students interfaced the hardware component with the RTOS part in an incremental approach. They first insured a communication in user-land before exploring the kernel code and integrating the driver.




\subsubsection{Integration into the Xenomai Nucleus}

All the code modifications and the integration of the hardware time manager calls were performed at the \textit{Nucleus} level. Doing so, we ensured the compatibility of the performed modifications with 1) all Xenomai skins and 2) kernel and user-land based Xenomai real-time applications. We included our driver into the Nucleus code, and modified the \verb+wait_period()+ primitive, which now calls the hardware time manager. This allows to launch a timer with an initial value corresponding to the task period, and then suspends the calling task.

When the delay is elapsed, the hardware component produces an interrupt signal. Thus, we implemented  a handler executed each time the interrupt is received. This function retrieves the identifier of the task(s) needing to be woke up. Then, it places it (them) in the ready state after sending an acknowledgment to the component indicating that the task is actually awaken.

\section{Performance evaluation}

In this section, after introducing the evaluation platform, we give some results describing the benefits gained from replacing the software time manager service by the hardware version. We implemented the hardware time manager on a development board and measured the time manager latencies. From these results we computed the corresponding CPU load for a given set of tasks. We also present the cost of the hardware component, in terms of FPGA resources. In the following section we refer to the software default Xenomai time manager as the \textit{software mode}, as opposed to the \textit{hardware mode}.

\subsection{Evaluation platform}

Our evaluations were achieved on the \textit{Armadeus Systems APF27} development board \cite{Armadeus12}. It is equipped with a \textit{Freescale i.MX27} microprocessor clocked at a frequency of 400 Mhz. This processor is coupled to a \textit{Xilinx Spartan 3A} FPGA chip on which we synthesized a small version of the hardware time manager. This component is able to manage up to 12 real-time tasks. The hardware time manager was clocked at a frequency of 102 MHz, giving one tick every 9.8 ns. For the software part we used the native Xenomai skin, in a \textbf{non-tickless mode}, with a base period of 10 ms. We used the user-land mode for the defined Xenomai real-time tasks.

\subsection{Performance evaluation results}
In this part, we investigate the performance of the proposed design throughout the reactivity, saved CPU load, and the FPGA resource cost.

\subsubsection{Reactivity and execution time}

\paragraph{Time measurement methodology}
In the next sections, we present results based on various time measurements. Those measurements were made, using the \verb+GetTime+ operation implemented by the hardware time manager as follows : 

\begin{verbatim}
/* 1. Measure the calibration time */
c1 = GetTime(); c2 = GetTime();
calibration_value = c2 - c1;
/* 2. Measure the execution time */
t1 = GetTime();
call_to_measured_operations();
t2 = GetTime();
result = (t2 - t1) - calibration_value;
\end{verbatim}

By subtracting the calibration value from the measured time, we took off the overhead due to the \verb+GetTime+ function itself (measured by two consecutive GetTime operations). Each time measure were performed 10 times and we considered the average value. The measures were always performed just after the system boot, thus insuring the same initial conditions.The hardware time manager bitstream were loaded at boot time (U-Boot) before the kernel starts up.

\begin{figure}
  \centering
  \includegraphics[width=0.5\textwidth]{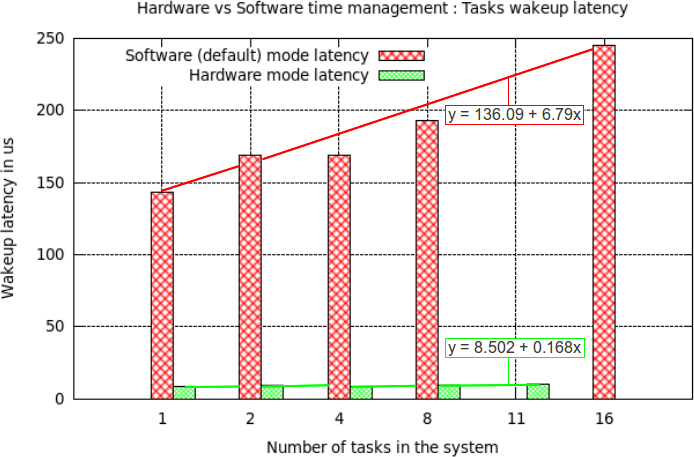}
  \caption{System reactivity, wake-up latency per task.}
  \label{Reactivity}
\end{figure}

\paragraph{System reactivity}
We investigate the system reactivity by quantifying the imprecision between the moment a task needs to be woke up (i.e. the occurrence of the tick corresponding to the end of its delay period) and the moment this task is really woke up. To do so, we measured the execution times of the tick management function in software mode, and compared them to the execution time of the task wake up interrupt handler in hardware mode. Results for different number of tasks are shown in Figure \ref{Reactivity}.

We can see that in software mode, the task wake-up latency is highly related to the number of tasks in the system. Indeed, at each clock cycle the master timer handler must inform all the delayed tasks timers that a clock cycle occurred. In the hardware mode, this latency is much less dependent on the number of tasks (see the estimated slope equation), thus  making it more stable. The inferred wake up latencies are highlighted in Figure \ref{Reactivity}.

\paragraph{Offloading the CPU}
Based on previously measured values (see  Figure \ref{Reactivity}), we can estimate the CPU overhead for delayed tasks management in both software and hardware modes for a duration of \textit{one second}. We assumed a master timer base period of 10 ms in the software mode. In order to simplify the figure, we also assumed that all tasks have the same period (no impact on the results). 

\begin{figure}
  \centering
  \includegraphics[width=0.5\textwidth]{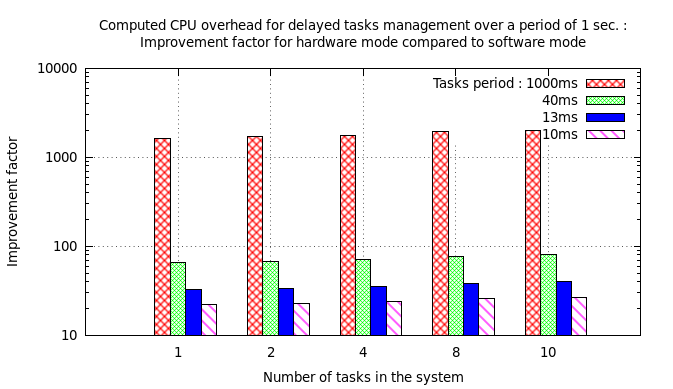}
  \caption{Improvement factor for the CPU overhead due to delayed tasks management. Improvement factor = (CPU overhead in software) / (CPU overhead in hardware)}
  \label{CPU_relief}
\end{figure}

In software mode, we computed the CPU time dedicated to the delayed tasks management using the estimation equation in Figure \ref{Reactivity} and multiplying the result by the number of clock ticks in one seconds : 100 (1 sec divided by 10 ms) in our case. In hardware mode, this time value corresponds to the number of task awakenings in a 1 second period multiplied by the time measured of the task wake up interrupt handler (in Figure \ref{Reactivity}). In Figure \ref{CPU_relief}, we present the improvement factor (speedup) given by using the hardware mode. This speedup is obtained by dividing the time for task delay management in software mode by the time taken in hardware mode. 

We can observe that the speedup is very high when the number of task wake-ups is low (i.e. the task period is long), because in hardware mode we spend CPU time only when needed (interrupt driven) in order to wake up tasks. Conversely, in software (polling) mode, we have to iterate over the delayed tasks timer list at each timer tick. One must keep in mind that for the hardware timer the precision is 3 orders of magnitude better than the software version for all the performed measures.

\subsubsection{FPGA hardware resource cost}

The FPGA resource cost is given in terms of FPGA 4 input LUTs, and flip-flop slices used by the hardware time manager on the Spartan3A chip. To obtain these values,  we lied upon the outputs of the Xilinx ISE design suite. We synthesized various versions of the component, each one able to manage a given number of tasks and we studied the variation of resource utilization. Even though the given implementation, students worked on, is very naive and can be substantially optimized, we found it interesting to show the results.

\begin{figure}
  \centering
  \includegraphics[width=0.5\textwidth]{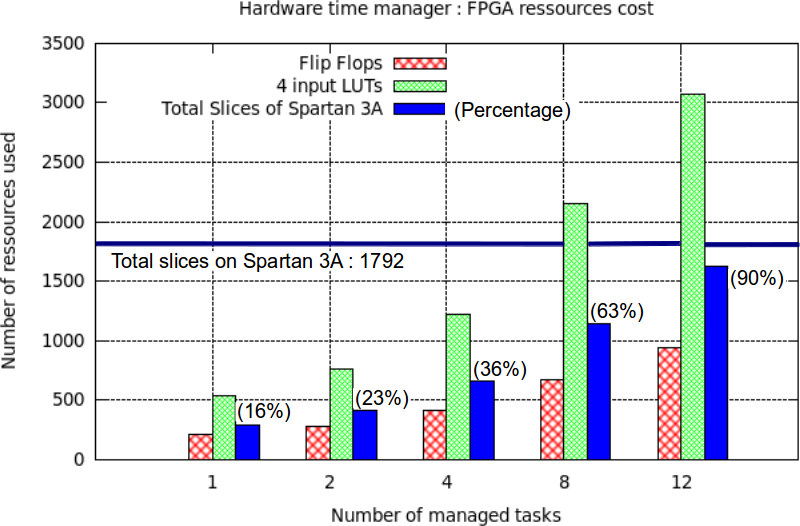}
  \caption{FPGA resources cost for various versions of the hardware timer according to the number of managed tasks}
  \label{FPGAcost}
\end{figure}

Results about FPGA resource costs are presented in Figure \ref{FPGAcost}. A maximum of 12 tasks managed on the Spartan3 may seem low, indeed we presented the timer with the higher granularity (64 bit counters). The cost in terms of hardware resources for the component can be optimized by reducing the granularity of the timer or optimizing the architecture.

\section{Conclusion and future works}
In this study we presented a simple architecture model for a flexible/configurable hardware time manager component, designed to replace the software version of the Xenomai kernel of embedded Linux. As our performance evaluation shows, this implementation drastically enhances the reactivity of the system by more than a factor  10, and allows to have a flexible and extremely precise timer for real-time tasks. Furthermore, we also considerably reduced the CPU overhead due to the delayed tasks management.This is not without a quantified FPGA hardware cost, but the flexibility and precision offered by such a component can be worth it. 
This work can be extended by testing the reconfigurability feature of some FPGAs (not the Spartan3A) to dynamically reconfigure the timer in terms of precision and number of managed tasks according to the application evolution. We plan to reduce the cost of the hardware component in terms of FPGA resources, in order to be able to manage a more important number of tasks

This project was given to Master degree students to introduce them to embedded systems research. It was well received by students and considered as challenging as it allowed them to apprehend many domains of embedded systems such as:  1) embedded Linux tools handling, 2) Xenomai real time kernel installation and use, 3) device driver development and integration, 4) kernel code exploration and programming, 5) hardware programming and interfacing, and 6) performance evaluation and validation. 

All the hardware VHDL and component driver sources together with a Xenomai patch will be available on-line.

\bibliographystyle{abbrv}
\bibliography{sigproc}

\begin{thebibliography}{1}

\bibitem{Armadeus12}
{Armadeus Systems}.
\newblock Apf27 board datasheet, 2012.
\newblock
  \url{http://www.armadeus.com/_downloads/apf27/documentation/datasheet_apf27.pdf}.

\bibitem{Burleson93}
W.~Burleson, J.~Ko, D.~Niehaus, K.~Ramamritham, J.~A. Stankovic, G.~Wallace,
  and C.~Weems.
\newblock The spring scheduling co-processor: A scheduling accelerator.
\newblock In {\em IEEE Transactions on VLSI}, 1993.

\bibitem{Gerum05}
P.~Gerum.
\newblock {\em Life with adeos}.
\newblock 2005.

\bibitem{Kiszka05}
J.~Kiszka.
\newblock The real-time driver model and first applications.
\newblock In {\em 7th {Real-Time} Linux Workshop, Lille, France}, 2005.

\bibitem{Kohout03}
P.~Kohout, B.~Ganesh, and B.~Jacob.
\newblock Hardware support for real-time operating systems.
\newblock In {\em Proceedings of the 1st IEEE/ACM/IFIP international conference
  on Hardware/software codesign and system synthesis}, CODES+ISSS '03, pages
  45--51, New York, NY, USA, 2003. ACM.

\bibitem{Kuacharoen03}
P.~Kuacharoen, M.~Shalan, and V.~M.
\newblock A configurable hardware scheduler for real-time systems.
\newblock In {\em in Proceedings of the International Conference on Engineering
  of Reconfigurable Systems and Algorithms}, pages 96--101. CSREA Press, 2003.

\bibitem{Labrosse02}
J.~Labrosse.
\newblock {\em {MicroC/OS-II:} the real-time kernel}.
\newblock Newnes, 2002.

\bibitem{Lindh91}
L.~Lindh.
\newblock Fastchart-a fast time deterministic cpu and hardware based
  real-time-kernel.
\newblock In {\em Real Time Systems, 1991. Proceedings., Euromicro '91 Workshop
  on}, pages 36 --40, jun 1991.

\bibitem{Yaghmour01}
K.~Yaghmour.
\newblock Adaptive domain environment for operating systems.
\newblock {\em Opersys inc}, 2001.

\end{thebibliography}

\balancecolumns

\end{document}